\documentclass[aps,pra,amsmath,amssymb,graphicx,superscriptaddress]{revtex4-2}
\usepackage{graphicx}
\usepackage{dcolumn}
\usepackage{bm}
\usepackage{amsmath}
\usepackage{amssymb}
\usepackage{physics}
\usepackage{nicefrac}
\usepackage[colorlinks=true, allcolors=blue]{hyperref}

\begin{document}

\title{History-Dependent Dynamical Invariants in the Lorenz System}

\author{B. A. Toledo}
\email{btoledoc@uchile.cl}
\affiliation{Departamento de Física, Facultad de Ciencias, Universidad de Chile, Santiago, Chile.}

\date{\today}

\begin{abstract}
Contrary to the established view of the Lorenz system as an archetype
of dissipative chaos lacking conserved quantities, this work
rigorously demonstrates the existence of a novel class of
history-dependent dynamical invariants. Through a constructive method
that augments the phase space, we derive a non-local invariant whose
value remains constant along any trajectory. Its history-dependence
arises from an integral term that accumulates the orbit's past,
thereby ensuring its conservation. The invariant's constancy is
verified with high-precision numerical simulations for both periodic
and chaotic orbits. This finding reveals a hidden structure within the
attractor and affords a new physical interpretation where unstable
periodic orbits (UPOs) correspond to specific values of this conserved
quantity. The result redefines the notion of non-integrability in
dissipative systems, showing that non-local order can coexist with
chaotic behavior.
\end{abstract}

\maketitle

\section{INTRODUCTION}
Conserved quantities, or constants of motion, are fundamental concepts
in the study of dynamical systems. They represent physical properties,
such as energy or momentum, that remain constant as a system
evolves. In Hamiltonian mechanics, Noether's theorem provides a
profound connection between the continuous symmetries of a system's
Lagrangian and its conserved quantities, offering deep physical
insights and often simplifying the analysis \cite{Goldstein2002}. The
existence of a sufficient number of independent conserved quantities
can render a system integrable, confining its motion to a torus in
phase space and leading to regular, predictable behavior. However,
many natural and engineered systems, particularly those exhibiting
complex or chaotic dynamics, are not conservative. This
non-conservative nature is typically associated with phase-space
volume contraction and the presumed absence of a sufficient number of
classical, time-independent conserved quantities required for
integrability. One of the most iconic examples of such a system is the
Lorenz system, introduced by Edward N. Lorenz in 1963 as a severely
truncated model of atmospheric Rayleigh-Bénard convection
\cite{Lorenz1963}. In the physical system of convection, an
equilibrium between energy input and dissipation can lead to sustained
periodic fluid motions, known as periodic orbits. Depending on the
system parameters, these orbits can be stable or unstable, with the
latter often preceding more complex, turbulence-like behavior. The
Lorenz model, designed to capture such transitions, is described by a
set of three coupled, nonlinear ordinary differential equations:
\begin{align}
\frac{dx}{dt}&=\sigma(y-x); \label{eq:lorenz1} \\
\frac{dy}{dt}&=x(\rho-z)-y;
\label{eq:lorenz2} \\
\frac{dz}{dt}&=xy-\beta z. \label{eq:lorenz3}
\end{align}
Here, $x$, $y$, and $z$ are state variables representing the intensity
of convective motion, the horizontal temperature difference, and the
distortion of the vertical temperature profile, respectively. The
parameters $\sigma$ (the Prandtl number), $\rho$ (the Rayleigh
number), and $\beta$ (a geometric factor) are positive constants. For
certain parameter values (e.g., Lorenz's classical values $\sigma=10$,
$\rho=28$, $\beta=8/3$), the system exhibits deterministic chaos,
characterized by a sensitive dependence on initial conditions and the
presence of a "strange attractor" \cite{Strogatz2015}. In
Eq.~\eqref{eq:lorenz1}, the fluid's acceleration ($\dot x$) arises
from the competition between the buoyancy force ($\sigma y$) driving
the flow and the viscous friction ($-\sigma x$) opposing the
motion. In Eq.~\eqref{eq:lorenz2}, the rate of change of the
horizontal temperature difference is governed by two competing
processes: it is generated by convective motion ($x(\rho-z)$) while
being simultaneously dissipated by thermal diffusion ($-y$). In
Eq.~\eqref{eq:lorenz3}, the distortion of the vertical temperature
profile is generated by the convective heat flux ($xy$) and is
dissipated by thermal diffusion ($-\beta z$), which acts to restore a
linear conductive profile.  It is crucial to distinguish the Lorenz
system from Hamiltonian systems. Hamiltonian systems, by definition,
conserve energy (if the Hamiltonian is time-independent) and preserve
phase-space volume, with dynamics governed by symplectic geometry. The
Lorenz system, being dissipative, does not fit this framework. Its
chaotic behavior arises from an intricate interplay of stretching
phase-space volumes, which leads to sensitivity to initial conditions,
and folding them, which keeps trajectories within a bounded region,
with an overall contraction of volume due to
dissipation. Consequently, the standard Lorenz system, particularly in
its chaotic regime, is generally understood to lack nontrivial,
time-independent conserved quantities. While there have been attempts
to find Lagrangian or Hamiltonian formulations for dissipative
systems, often employing fractional calculus \cite{Tarasov2010,
  Cresson2007} or embedding them within higher-dimensional
conservative frameworks, these approaches typically yield conserved
quantities for modified systems and do not straightforwardly apply to
the original three-dimensional dissipative equations without careful
reinterpretation.  This paper challenges this conventional
understanding by investigating the existence and nature of dynamical
invariants with a history-dependent character within the standard
Lorenz equations \eqref{eq:lorenz1}-\eqref{eq:lorenz3}. We aim to
demonstrate that, despite its dissipative and chaotic nature, the
Lorenz system possesses such history-dependent dynamical
invariants. The analysis will begin by examining the fundamental
dynamical properties of the Lorenz system, emphasizing its dissipative
character and known symmetries. Subsequently, we will present the
derivation of these invariants and explore their implications, placing
them in the context of generalized conservation laws.  The
significance of the Lorenz system extends far beyond its
meteorological origins, serving as a paradigm for chaotic behavior
with profound implications across numerous scientific and engineering
disciplines. Its relevance often stems precisely from its
deterministic chaos, sensitive dependence on initial conditions, and
its dissipative, non-integrable nature. In atmospheric science and
fluid dynamics, it remains a key model for understanding the limits of
predictability in weather and for basic thermal convection
\cite{Palmer1993}, with extensions modeling complex geophysical
phenomena. Applications in physics include laser instabilities
\cite{Haken1975}, plasma dynamics, and dynamo theory. In engineering,
it has been utilized in nonlinear circuit theory, chaos control
\cite{Pecora1990}, and secure communications \cite{Liu2022}. It finds
analogies in chemical reactions where transient chaotic phenomena are
observed \cite{Scott1991}, and informs models in biology
\cite{Schwartz2008}, medicine, and economics
\cite{Mandelbrot1963}. Its role in mathematics is also central,
particularly in studies of strange attractors and bifurcations
\cite{Ghys2007, BirmanWilliams1983}. Its broad applicability arises
from its qualitative capture of deterministic chaos from simple
nonlinear interactions. The very characteristics that traditionally
challenge notions of integrability, including the subtleties
surrounding its conserved quantities, are what make it a universal
prototype. This paper seeks to provide a comprehensive explanation of
how the concept of dynamical invariants applies to the Lorenz system,
thereby reshaping our understanding of its complex behavior.

\section{THE LORENZ SYSTEM: DYNAMICS AND DISSIPATION}
The Lorenz equations \eqref{eq:lorenz1}-\eqref{eq:lorenz3} define a
continuous-time dynamical system in a three-dimensional phase
space. The behavior of this system is critically dependent on the
values of the parameters $\sigma$, $\rho$, and $\beta$.
\subsection{Dissipative Nature}
A key characteristic of the Lorenz system is its dissipative
nature. The divergence of the vector field
$F=(\sigma(y-x),x(\rho-z)-y,xy-\beta z)$ is given by:
\begin{align}
\nabla\cdot F&=\frac{\partial}{\partial x}(\sigma(y-x))+\frac{\partial}{\partial y}(x(\rho-z)-y)+\frac{\partial}{\partial z}(xy-\beta z) \nonumber \\
&=-\sigma-1-\beta.
\label{eq:divergence}
\end{align}
Since $\sigma, \beta > 0$, the divergence is a negative constant. This
implies that any volume element in phase space contracts exponentially
with time, its volume scaling as $e^{-(\sigma+1+\beta)t}$
\cite{Ott2002}. This volume contraction is a hallmark of dissipative
systems. It means that, as time progresses, trajectories are attracted
to a subset of the phase space with zero volume. In the chaotic
regime, this subset is a fractal strange attractor \cite{Alberti2023};
in non-chaotic regimes, it may consist of a finite set of points
\cite{kuznetsov2020lorenz}.
\subsection{Fixed Points and Stability}
The fixed points of the Lorenz system are found by setting
$dx/dt=dy/dt=dz/dt=0$. One fixed point is always at the origin
$(0,0,0)$, corresponding to a state of no convection. If $\rho>1$, two
additional fixed points, $C^{+}$ and $C^{-}$, emerge:
\begin{equation}
(x,y,z)=(\pm\sqrt{\beta(\rho-1)},\pm\sqrt{\beta(\rho-1)},\rho-1). \label{eq:fixedpoints}
\end{equation}
These points represent steady convection. The stability of these fixed
points depends on the system parameters. The origin is stable for
$\rho<1$. For $\rho>1$, the origin becomes unstable, and $C^{+}$ and
$C^{-}$ become stable. However, as $\rho$ increases further, $C^{+}$
and $C^{-}$ also lose stability via a subcritical Hopf bifurcation at
$\rho_{H}=\sigma(\sigma+\beta+3)/(\sigma-\beta-1)$, provided
$\sigma>\beta+1$ \cite{Sparrow1982}. For Lorenz's parameters, this
occurs at $\rho\approx24.74$.

\subsection{The Strange Attractor}
For $\rho>\rho_{H}$ (e.g., $\rho=28$), trajectories do not settle into
a fixed point or a simple periodic orbit. Instead, they are confined
to a complex, bounded region in phase space known as the Lorenz
attractor. This attractor has a fractal structure and is the geometric
manifestation of the system's chaotic dynamics. Trajectories on the
attractor exhibit a sensitive dependence on initial conditions: nearby
trajectories diverge exponentially, making long-term prediction
impossible despite the deterministic nature of the equations. This
"butterfly effect" was one of Lorenz's key discoveries
\cite{Lorenz1963}.

\subsection{Symmetries of the Lorenz System}
The Lorenz system exhibits a discrete symmetry corresponding to the
transformation $(x,y,z)\mapsto(-x,-y,z)$, which leaves the
differential equations invariant. This $\mathbb{Z}_{2}$ symmetry
implies that the system is equivariant under a rotation of $180^\circ$
about the z-axis. As a result, the phase-space structure, including
fixed points and trajectories, is symmetric under this rotation. This
symmetry explains the appearance of the twin-lobed Lorenz attractor
and the existence of the symmetric pair of nontrivial fixed points,
$C^{+}$ and $C^{-}$. Although this discrete symmetry does not yield a
conserved quantity via Noether's theorem, which applies to continuous
symmetries of a Lagrangian, it plays a crucial role in constraining
the geometry and bifurcation structure of the system. The
$\mathbb{Z}_{2}$ symmetry also affects the organization of unstable
manifolds and the transition pathways between different regions of the
attractor, making it a fundamental structural feature of the Lorenz
dynamics.

\section{A CONSTRUCTIVE METHOD FOR HISTORY-DEPENDENT INVARIANTS}
In classical mechanics, a conserved quantity, or first integral,
$C(\mathbf{x})$, where $\mathbf{x}\in\mathbb{R}^{n}$, is a function
whose value remains constant along any trajectory of the system, i.e.,
$dC/dt=0$. If $C$ does not explicitly depend on time, it is a
time-independent conserved quantity. This section introduces a
distinct class of dynamical invariants for the Lorenz system that are
history-dependent.
\subsection{Derivation of the Dynamical Invariant $C_1$}
For the standard Lorenz system in its chaotic regime, it has long been
understood that no nontrivial, time-independent, analytic constants of
motion exist \cite{Sparrow1982, Tabor1989}. The dissipative nature of
the system is a strong indicator against the existence of classical
conserved quantities akin to energy in Hamiltonian systems, where
Liouville's theorem guarantees the preservation of phase-space
volume. The following analysis employs a modified direct method to
search for dynamical invariants.  To illustrate the approach, let us
consider a simple harmonic oscillator, $\ddot x + \omega^2 x = 0$,
which can be written as a first-order system:
\begin{align*}
  \dot x = y,\quad \dot y = -\omega^2 x.
\end{align*}
A conserved quantity $C(x,y)$ must satisfy $dC/dt = \nabla C \cdot (\dot x, \dot y) = \nabla C \cdot (y, -\omega^2 x) = 0$. This condition implies that $\nabla C$ is orthogonal to the flow vector $(y, -\omega^2 x)$. A simple choice for the gradient is $\nabla C = (\omega^2 x, y)$, which is proportional to $(-\dot y, \dot x)$. Integrating the components yields
\begin{align*}
  \frac{\partial C}{\partial x} = \omega^2 x \quad &\Rightarrow \quad C(x,y) = \frac{1}{2}\omega^2 x^2 + c_1(y), \\
  \frac{\partial C}{\partial y} = y \quad &\Rightarrow \quad C(x,y) = \frac{1}{2}y^2 + c_2(x).
\end{align*}
Reconciling these two expressions by setting $c_1(y) = \frac{1}{2}y^2$
and $c_2(x) = \frac{1}{2}\omega^2 x^2$ yields the total energy,
$C(x,y) = \frac{1}{2}\omega^2 x^2 + \frac{1}{2}y^2 \equiv E$. An
equivalent procedure, which is analogous to the method used for the
Lorenz system, involves summing the two partial results to obtain $2C
= \nicefrac{1}{2}\,\omega^2 x^2 + \nicefrac{1}{2}\,y^2$, by setting
$c_1(y)=c_2(x)=0$. Since any function of a conserved quantity is also
conserved, we may, without loss of generality, rescale the invariant
by a constant factor. Redefining $C \to C/2$ yields the canonical form
of the energy.

This procedure can be generalized. For a system in an even-dimensional
phase space ($2n$), one can check all combinations leading to
orthogonality. If the dimension is odd ($2n+1$), as it is for the
Lorenz system, we can augment the space with an auxiliary variable
$u$, whose dynamics will be determined throughout the calculation. For
the Lorenz system, we seek an invariant $C(x,y,z,u)$. The conservation
condition is
\begin{align*}
  0 = \frac{dC(x,y,z,u)}{dt} = \nabla C \cdot (\dot x, \dot y, \dot z, \dot u),
\end{align*}
where $\nabla \equiv (\partial_x, \partial_y, \partial_z,
\partial_u)$. We let the dynamics of the auxiliary variable, $\dot u =
f(x,y,z,u)$, be determined by the consistency of the method. One
possible choice for the orthogonal vector is
\begin{equation}
\nabla C = (-\dot y, -\dot u, \dot z, -\dot x).
\end{equation}
Integrating each component yields four partial expressions for $C$:
\begin{align*}
  C(x,y,z,u) &= \frac{1}{2}x^2(z-\rho) + xy + c_1(y,z,u),\\
  C(x,y,z,u) &= -\int \dot u dy + c_2(x,z,u),\\
  C(x,y,z,u) &= xyz - \frac{1}{2}\beta z^2 + c_3(x,y,u),\\
  C(x,y,z,u) &= \sigma(x-y)u + c_4(x,y,z).
\end{align*}
The functions $c_k$ provide the necessary degrees of freedom to ensure
that these four expressions can be reconciled into a single,
self-consistent function. The ansatz for $\nabla C$ is not unique;
other choices are possible. Furthermore, the term $\partial_y C =
-\dot{u}$ suggests that $C$ might need to depend on velocities. To
maintain the assumption that $C$ is a function of state variables
only, we use the flexibility of the functions $c_k$ to absorb any
velocity dependence. This can be formalized by promoting the $c_k$ to
functions $\bar{c}_k$ that depend on velocities and then imposing a
simplifying condition, such as $\bar{c}_1 + \bar{c}_2 + \bar{c}_3 +
\bar{c}_4 + \int \dot{u} dy = 0$. This procedure forces consistency
and shifts the velocity dependence into a governing equation for $u$.
Combining the partial expressions and imposing this consistency leads
to a candidate for the invariant (after re-scaling $C \to C/4$):
\[ C(x,y,z,u) = xy(z+1) + \frac{1}{2}x^2(z-\rho) - \frac{1}{2}\beta z^2 + \sigma(x-y)u.
\]
The condition $dC/dt=0$ is now imposed, which yields a consistency
equation for the dynamics of $u$ in the form of a first-order linear
ordinary differential equation:
\begin{align*}
  \dot u &= F(x,y,z)u + G(x,y,z),
\end{align*}
where
\begin{align*}
F(x,y,z) &= \sigma - \frac{y+x(z-\rho)}{x-y},\\
G(x,y,z) &= \frac{1}{2\sigma(x-y)}\left[x^{2}(z(\beta+2(z+1+\sigma-\rho))-2\rho(\sigma+1)-2y^{2})-x^{3}y+2xy((2\beta+1)z+(\rho+1)\sigma+1)\,+\right.\\
  &\phantom{1111111111111}\left.-2(\sigma y^{2}(z+1)+\beta^{2}z^{2})\right].
\end{align*}
Solving for $u(t)$ with an integrating factor gives:
\begin{equation*}
u(t)=\exp\left(\int_{0}^{t}\Phi_{1}(\tau_{1})d\tau_{1}\right) \left[ u(0) + \int_{0}^{t}\exp\left(-\int_{0}^{\tau_{2}}\Phi_{1}(\tau_{1})d\tau_{1}\right)\Phi_{2}(\tau_{2})d\tau_{2} \right],
\end{equation*}
where $\Phi_{1}(t)=F(x(t),y(t),z(t))$ and $\Phi_{2}(t)=G(x(t),y(t),z(t))$.
Substituting this into the expression for $C$ yields the first history-dependent invariant, $C_1$:
\begin{equation}
C_{1}(t) = xy(z+1) + \frac{1}{2}x^2(z-\rho) - \frac{\beta z^2}{2} + \sigma(x-y)u(t).
\label{eq:C1time}
\end{equation}
This quantity is conserved by construction. Its value depends on the
instantaneous state $(x(t), y(t), z(t))$ and on the entire history of
the trajectory, which is embedded within the integral terms defining
$u(t)$. However, we readily note that both $F$ and $G$ are undefined
at $x=y$. To resolve this issue, we consider the variable
$v=(x-y)u$. It then follows that
\begin{align*}
  \dot v &= (\dot x-\dot y)u + (x-y)\dot u\\
  &= \left[\sigma(y-x) - (x(\rho-z)-y)\right]u + (x-y)\left[F(x,y,z)u+G(x,y,z)\right]\\
  &= \left[\sigma(y-x) - x(\rho-z)+y\right]u + (x-y)\left[\left(\sigma - \frac{y+x(z-\rho)}{x-y}\right)u+G(x,y,z)\right]\\
  &= (x-y)G(x,y,z)\\
  &= \frac{1}{2\sigma}\left[x^{2}(z(\beta+2(z+1+\sigma-\rho))-2\rho(\sigma+1)-2y^{2})-x^{3}y+2xy((2\beta+1)z+(\rho+1)\sigma+1)\,+\right.\\
    &\phantom{1111111111111}\left.-2(\sigma y^{2}(z+1)+\beta^{2}z^{2})\right],\\
  v(t)&=\frac{1}{2\sigma}\int_0^t\left[x^{2}(\tau)(z(\tau)(\beta+2(z(\tau)+1+\sigma-\rho))+\ldots\right]\,d\tau,
\end{align*}
where the singularity has been removed. Therefore, the regularized form of $C_1$ is
\begin{subequations}\label{eq:C1Rtime}
\begin{align}
  C_{1}(t) &= xy(z+1) + \frac{1}{2}x^2(z-\rho) - \frac{\beta z^2}{2} + \sigma v(t), \label{eq:C1Rtime-a} \\
  &= xy(z+1) + \frac{1}{2}x^2(z-\rho) - \frac{\beta z^2}{2} + \frac{1}{2}\int_0^t\left[x^{2}(\tau)(z(\tau)(\beta+2(z(\tau)+1+\sigma-\rho))+\ldots\right]\,d\tau.
\label{eq:C1Rtime-b}
\end{align}
\end{subequations}
It is important to note that if we take $u(0)=0$, both the regularized
and non-regularized versions of $C_1$ respect the $\mathbb{Z}_2$
symmetry; therefore, we fix $u(0)=0$. Also note that
Eqs.~(\ref{eq:C1Rtime-a}) and (\ref{eq:C1Rtime-b}) are analytically
equivalent; however, Eq.~(\ref{eq:C1Rtime-a}) is faster, and more
accurate, to evaluate with high numerical precision. Therefore, we
will use Eq.~(\ref{eq:C1Rtime-a}) in numerical calculations, which
implies solving the extended system.  The invariant, $C_1$, is a
conserved functional whose mathematical structure warrants detailed
analysis. Its components, such as $\frac{1}{2}x^{2}(z-\rho)$ and
$-\frac{\beta z^{2}}{2}$, are formally analogous to terms of potential
and kinetic energy derived from the physical model of Rayleigh-Bénard
convection~\cite{Lorenz1963, Strogatz2015}. Nevertheless, the crucial
component ensuring conservation is the integral term $\sigma
v(t)$. This term, which depends on the complete history of the
trajectory, has no direct analogue in classical conservation
laws. Rather than acting as an "energy reservoir," the regularized
variable $v(t)$ functions as a memory component or a non-local
adjustment term, dynamically compensating for fluctuations in the
local terms to maintain the total value of $C_1$ strictly constant
along any given trajectory. This use of auxiliary variables to
establish conservation laws is a known feature in the formulation of
non-conservative systems \cite{Riewe1996, Musielak2008}.
\subsection{Invariant Structure and Chaotic Dynamics: The UPO-Eddy Analogy}
We now connect this mathematical structure to the physical intuition
of turbulence. The Lorenz system is a prototypical model for the
transition to chaos. In the real phenomena it models, energy injected
at a macroscopic scale cascades to smaller scales through a hierarchy
of eddies, a process conceptualized by Richardson and formalized by
Kolmogorov. While the Lorenz system lacks this spatial cascade, it
effectively captures the intrinsic flow instabilities that are the
fundamental mechanism of chaos. This chaos is structured by a
"skeleton" of an infinite number of unstable periodic orbits (UPOs)
\cite{Auerbach1987}. We propose the following heuristic analogy: the
UPOs are the dynamical representation of transient, coherent
structures—eddies—in a turbulent flow. A chaotic trajectory is an
unpredictable journey between these UPOs. From the phenomenology of
dynamical systems, we know that periodic orbits change their stability
through bifurcations. A transition from a stable to an unstable orbit
can be seen as the dynamical analogue of a large eddy splitting into
smaller eddies in a turbulent cascade. In this perspective, we propose
that the invariant $C_1$ could be related to the persistence of this
chaotic skeleton. Each UPO, or "bone" of the structure, corresponds to
a specific, constant value of the invariant. This interpretation
reframes the numerical challenge of tracking UPOs: their inherent
instability is not a methodological flaw but a feature, reflecting, as
a distant echo, the physical reality that eddies are intrinsically
unstable and ephemeral components of the turbulent
cascade~\cite{Moon2019,Park2021}.
\subsection{Numerical Tests on Periodic and Chaotic Orbits}

\begin{figure*}[ht]
\includegraphics[width=0.45\textwidth]{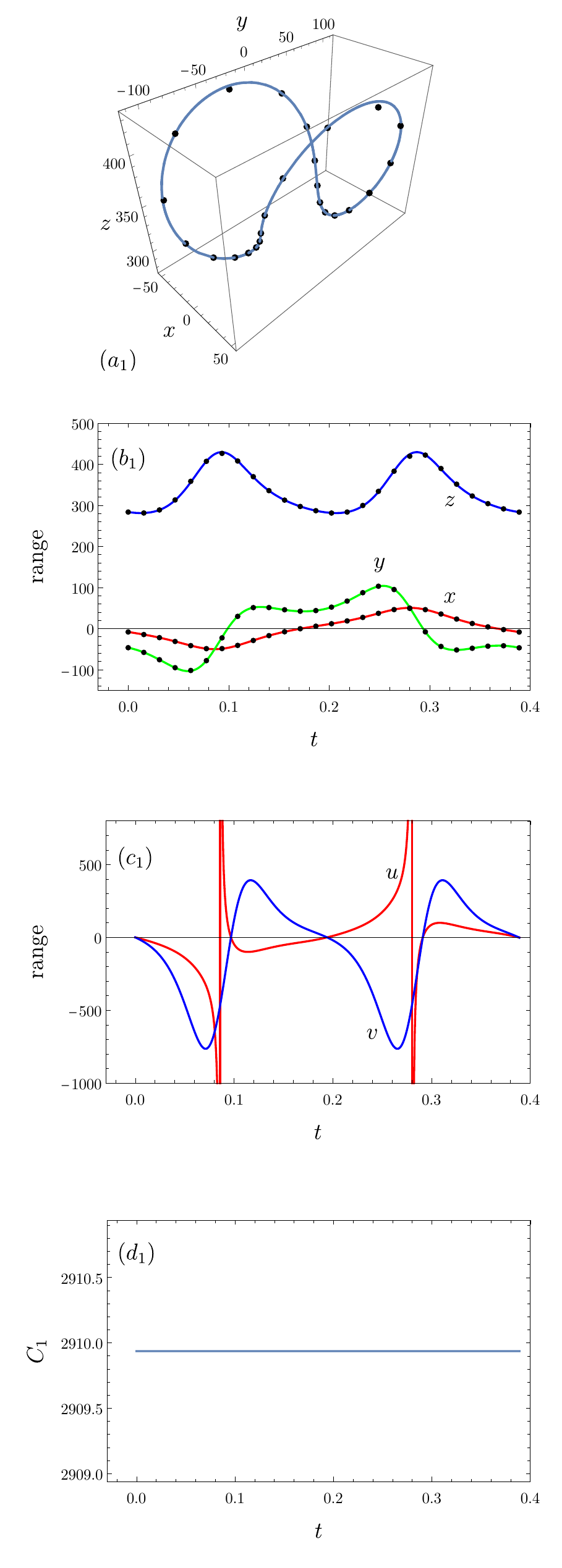}~\includegraphics[width=0.45\textwidth]{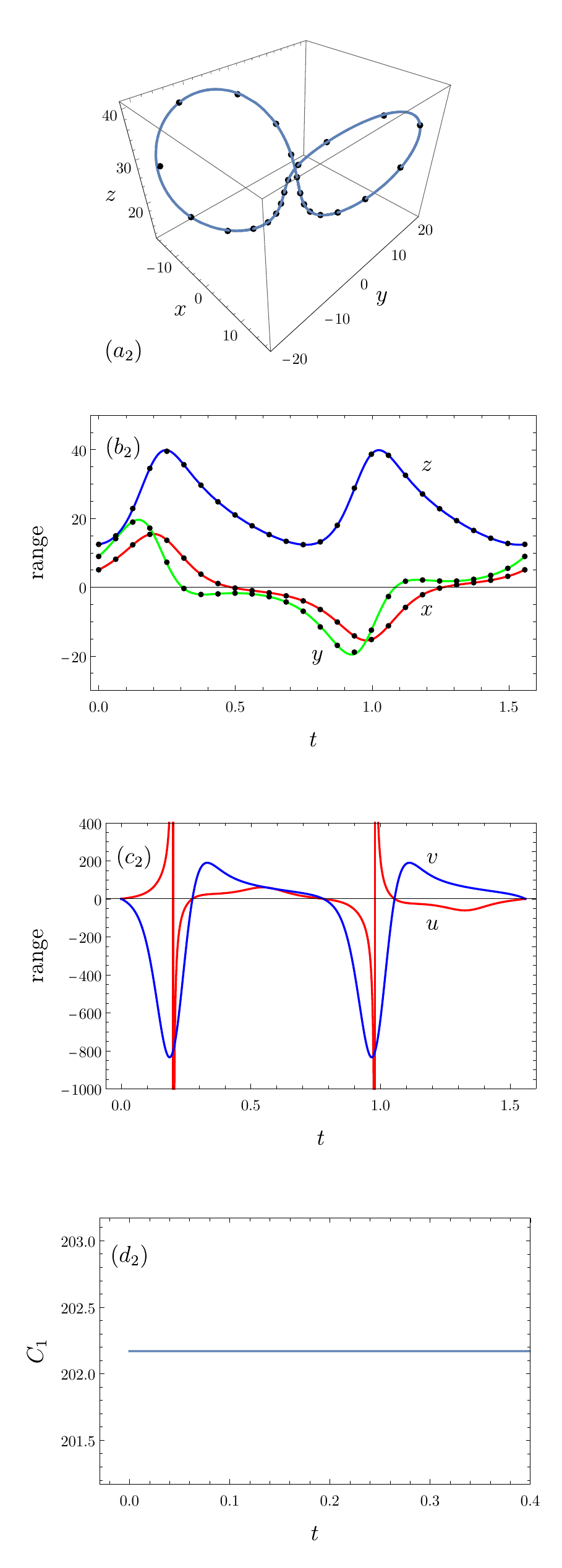}
\caption{The column with subindex 1 shows a stable periodic orbit for
  $\sigma=10, \rho=350, \beta=8/3$, displaying the time series of $x,
  y, z,$ and also of $u$ and $v$. The column with subindex 2 shows the
  same analysis for an unstable periodic orbit (UPO) with the
  classical parameters $\sigma=10, \rho=28, \beta=8/3$. This
  calculation was performed using multiprecision arithmetic at 150
  digits for the continuous lines, while the data represented by dots
  were calculated at machine precision.}\label{fig:figure1}
\end{figure*}

\begin{figure*}[ht]
\includegraphics[width=0.45\textwidth]{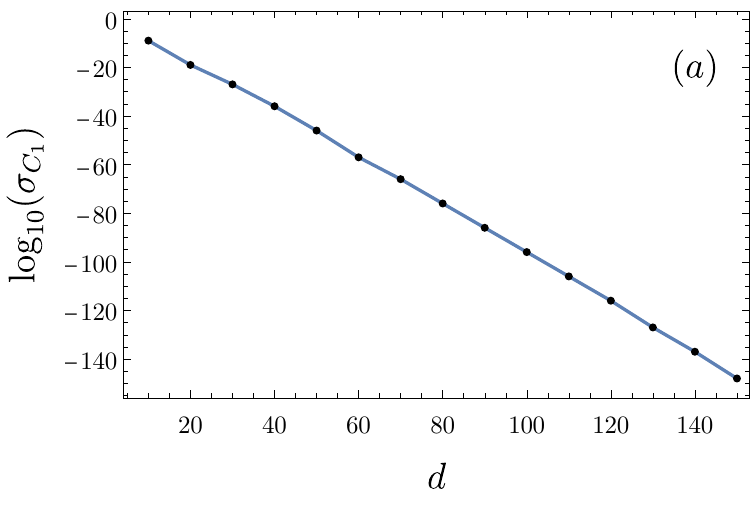}~\includegraphics[width=0.45\textwidth]{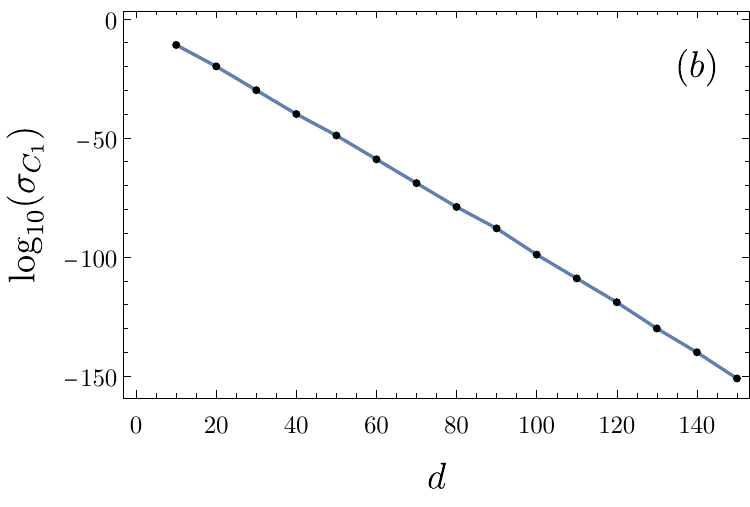}
\caption{Standard deviation of the regularized form of $C_1$ as a
  function of the number of digits of precision. Panel ($a$)
  corresponds to the stable orbit ($\sigma=10, \rho=350, \beta=8/3$),
  and panel ($b$) shows the case for the unstable orbit ($\sigma=10,
  \rho=28, \beta=8/3$). In each case the system was integrated for its
  corresponding period.}\label{fig:figure2}
\end{figure*}

\begin{figure*}[ht]
  \includegraphics[width=0.35\textwidth]{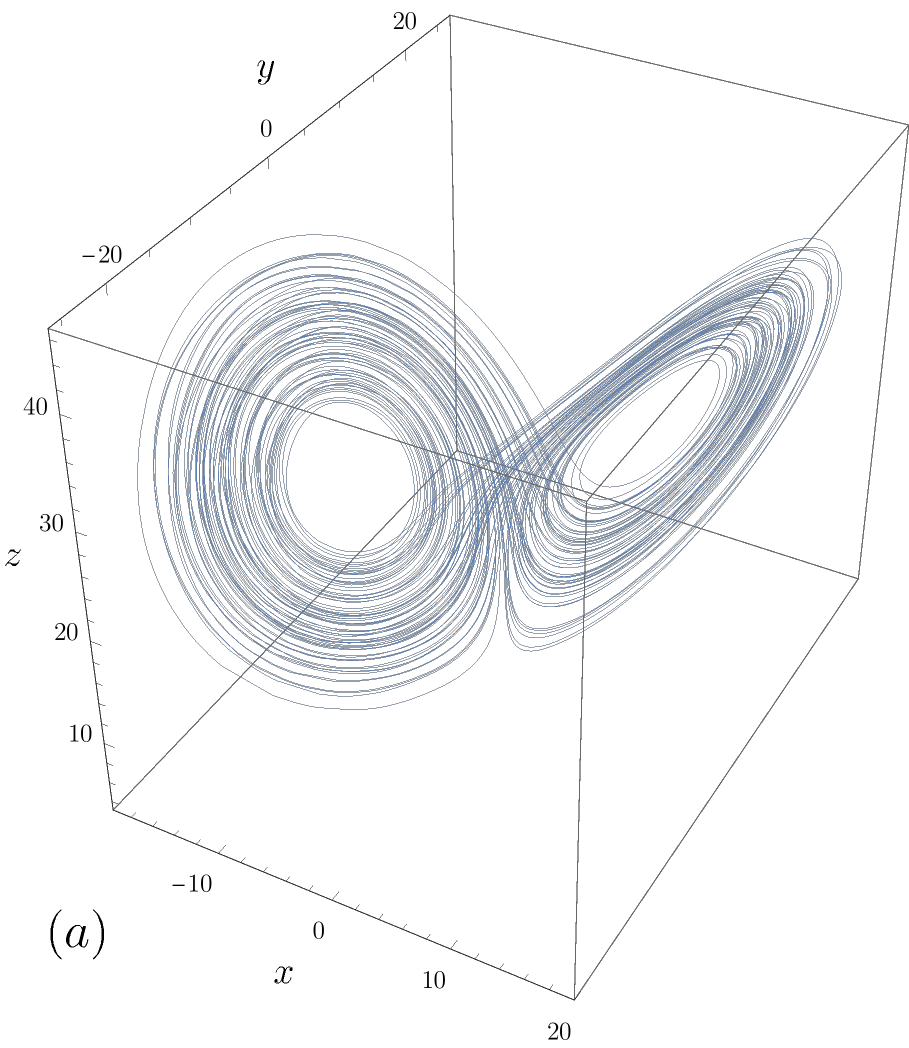}
  \includegraphics[width=0.5\textwidth]{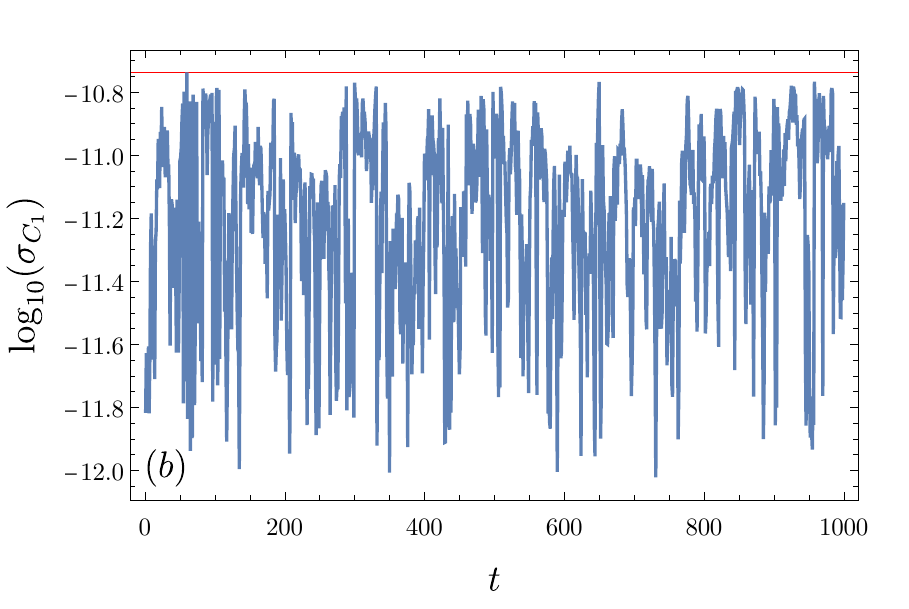}~\includegraphics[width=0.5\textwidth]{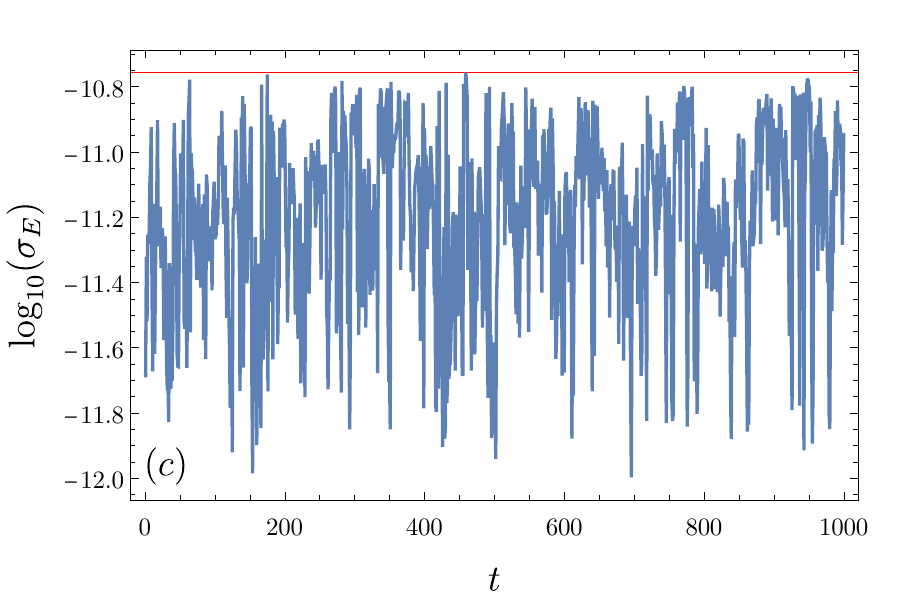}
\caption{A trajectory starting at a random point
  ($x_0=-16.164\ldots,\, y_0=-10.847\ldots,\, z_0=42.088\ldots$) on
  the attractor is shown in the upper figure. For the lower figures,
  ($b$), shows the dispersion of $C_1$ for an average value of
  $7033.8699495\ldots\;$ as evaluated along the trajectory (note the
  reference line for the maximal dispersion value). Here we use
  $\sigma=10, \rho=28, \beta=8/3$. On the right ($c$), for comparison,
  is shown the same calculation for the harmonic oscillator, with
  average energy $E=7000$. The calculations to evaluate $C_1$ and $E$,
  and their respective dispersions, used 50 digits of
  precision.}\label{fig:figure3}
\end{figure*}

To test this hypothesis, we focus on two specific periodic orbits
within the extended Lorenz system, and a chaotic one. The first, an
unstable periodic orbit (UPO), is defined by the initial conditions
$x(0)=5.05105\ldots$, $y(0)=8.93841\ldots$, $z(0)=12.44829\ldots$, and
$v(0)=0.0$, with a period of $\tau=1.55865\ldots$. The selection of a
specific UPO does not compromise generality, as Periodic Orbit Theory
(POT) posits that the system's asymptotic behavior is encoded in the
ensemble of all UPOs \cite{Cvitanovic2005}. The regularized, extended
four-dimensional system was numerically integrated using the
``StiffnessSwitching'' method in {\it Mathematica} to manage stiffness
while advancing along the orbit. Figure~\ref{fig:figure1} displays the
results for this UPO and for a second, stable periodic orbit at
$\rho=350$ with a period of $\tau=0.388\ldots$, using the same values
for $\sigma$ and $\beta$; in
Figs~\ref{fig:figure1}($a_1,\,a_2,b_1,b_2$) we calculate the solutions
with very high precision ($d=150$), this corresponds to continuous
lines, while dots represent the solution at machine precision; in
Figs~\ref{fig:figure1}($c_1,\,c_2$), we show the auxiliary variables,
note that functions in Figs~\ref{fig:figure1}($c_1$), were rescaled
for better comparison in the same plot ($u\to u/10,\,v\to v/200$);
finally, in Figs~\ref{fig:figure1}($d_1,\,d_2$), we show the value of
$C_1$ along the trajectory. In both cases, as shown in
Fig.~\ref{fig:figure2}, the dispersion of the invariant decreases
monotonically with increasing precision, providing strong numerical
evidence that $C_1$ is indeed a constant of motion, here we took
special care of not using interpolating function in the internal
process of calculations, since it diminishes the final precision,
Fig.~\ref{fig:figure2} corresponds to the most accurate calculation
performed for this study. Finally, as depicted in
Fig.~\ref{fig:figure3}, we selected a random point on the
attractor—defined by $x_0=-16.164\ldots$, $y_0=-10.847\ldots$, and
$z_0=42.088\ldots$—using the standard chaotic parameter values. The
system was then integrated over 1000 time units in intervals of
$\Delta t=1$; within each interval, the integration proceeded as in
the previous examples (using interpolation for a dense output,
therefore less local precision). At the conclusion of each time
interval, we calculated the value of $C_1$ and its standard deviation,
$\sigma_{C_1}$, from a sample of 100 equally spaced points within that
interval. The final state of each interval served as the initial
condition for the subsequent one. For comparison, we simulated the
harmonic oscillator for
$\omega=1,\,x_0=\sqrt{7000},\,y_0=\sqrt{7000}$, using exactly the same
procedure. It is notable how the dispersion of both conserved
quantities exhibits the same behavior.

\section{DISCUSSION AND FUTURE DIRECTIONS}
This work establishes the existence of nontrivial history-dependent
dynamical invariants in the Lorenz system. This finding holds
throughout the chaotic attractor, where each periodic orbit is
associated with a specific, constant value of the invariant. A key
aspect of this invariant is its non-local character in time, imparted
by an integral component. This feature marks a conceptual departure
from classical conserved quantities arising from Noether’s theorem,
which are local functions of the system’s instantaneous state
\cite{Tabor1989, Noether1918}. This history-dependence could be
interpreted physically as representing memory effects within the
turbulent cascade or non-local correlations in the flow field not
captured by local variables alone.  At this point, the nature of the
auxiliary variable $u$ is purely mathematical: it is a tool for
defining the invariant, and it is not yet clear whether it represents
a true physical quantity. However, one can recall the electromagnetic
vector potential, $\mathbf{A}(\mathbf{r},t)$, which is not directly
observable but is indispensable for a consistent formulation of the
theory. The variable $u$ could be seen as a necessary mathematical
degree of freedom, and setting $u(0)=0$ as a gauge fixing. In any
case, a real physical quantity that behaves like $u$, or its
regularized form $v$, should be sought in the full physical system
from which the Lorenz system is derived.  The implications of
rigorously establishing this invariant are significant. While this
study has focused on the derivation and analysis of one such
invariant, the constructive method employed can yield other conserved
quantities. A full, systematic exploration of the complete set of
invariants is deferred to a future study. The present work serves to
establish the principle that the Lorenz system, and likely other
dissipative systems, may be endowed with a richer structure of
dynamical constraints than previously understood. This opens several
avenues for future work:

\begin{itemize}
\item \textbf{Theoretical}: It is essential to investigate whether
  these invariants are linked to underlying, non-obvious symmetries. A
  systematic exploration could reveal whether the differing symmetry
  properties of other invariants correspond to distinct physical
  quantities, such as energy (symmetric) versus helicity
  (chiral). Furthermore, these renewed tools could offer new insights
  into classical problems that are now resolved, such as Smale's 14th
  problem~\cite{Tucker2002}. Although the existence of the Lorenz
  attractor has been rigorously demonstrated, these methods could
  enable a deeper quantitative analysis of its structure or of related
  chaotic systems.
\item \textbf{Computational}: The confirmation of these invariants
  offers a crucial validation target for emerging data-driven
  methodologies, including symbolic regression \cite{Angelis2023,
    Schmidt2009} and machine learning algorithms \cite{Cava2021}
  designed to uncover hidden invariants from time-series data.
\item \textbf{Physical}: The role of these invariants in networks of
  coupled Lorenz-type oscillators
  \cite{Krishnamurthy1999}—particularly in the emergence of collective
  behavior and synchronization—remains an open and compelling area for
  investigation.
\end{itemize}
Ultimately, the revelation that this complex behavior can coexist with
the internal constraints of these history-dependent invariants could
reshape our understanding of chaos itself, suggesting once more that
even within dissipative and turbulent regimes, order and structure may
persist~\cite{Miranda2013}.

\section{CONCLUSION}
The Lorenz system, a pillar of nonlinear dynamics, has traditionally
been characterized by the absence of classical, time-independent
conserved quantities within its chaotic regime. This view, linked to
its dissipative nature, has supported the interpretation of the system
as fundamentally non-integrable in the sense of Liouville-Arnold. This
work, while not refuting the chaotic and non-integrable nature of the
system, necessitates a significant revision of what non-integrability
implies. We demonstrate that the absence of \textit{classical} first
integrals is not equivalent to a total absence of structure. Instead,
we reveal the existence of a form of hidden, non-local order, namely a
history-dependent integral constraint, that coexists with chaos,
redefining our understanding of the interplay between structure and
complexity in dissipative systems. This is particularly true for
periodic orbits, regardless of their stability, since quantities like
$C_1$ allow them to be defined as parametric curves in phase space
with a well-defined mathematical form. These invariants demonstrate
that, contrary to conventional wisdom, constraining structures can
coexist with dissipation and chaos. While the system remains
non-Hamiltonian, the existence of these invariants reconfigures our
understanding of its internal dynamics. The system’s well-known
$\mathbb{Z}_2$ symmetry can now be understood in the context of a
richer structure that gives rise to these invariants, suggesting a
deeper geometric or algebraic underpinning. Rather than diminishing
the Lorenz system’s importance as a model of chaotic behavior, this
newfound structure enhances its value as a paradigm for studying the
interplay between order and chaos.
\begin{acknowledgments}
  I would like to acknowledge the invaluable contributions of the
  anonymous referees. Their critical perspectives and constructive
  suggestions have significantly strengthened the quality and
  presentation of this study. The author conducted this research
  independently at the Departamento de Física, Facultad de Ciencias,
  Universidad de Chile, without external funding.
\end{acknowledgments}

\section*{DATA AVAILABILITY STATEMENT}
Numerical calculations were performed using Mathematica, employing the
\texttt{NDSolve} function. The primary solver was configured with
\texttt{Method -> ``StiffnessSwitching''}, an adaptive meta-method
designed to efficiently integrate systems that exhibit both stiff and
non-stiff characteristics. This strategy initiates integration with a
non-stiff solver, which by default is an extrapolation method based on
the \texttt{ExplicitModifiedMidpoint} rule. The solver periodically
performs a stiffness test based on estimating the dominant eigenvalue
of the local Jacobian. If stiffness is detected, the method
automatically switches to a solver suitable for stiff problems, by
default an extrapolation method based on the
\texttt{LinearlyImplicitEuler} rule. This automated switching
mechanism ensures both computational efficiency in non-stiff regions
and numerical stability where the system becomes stiff. Other options
included \texttt{MaxSteps -> Infinity} to prevent premature
termination, and variable \texttt{WorkingPrecision},
\texttt{AccuracyGoal}, and \texttt{PrecisionGoal} settings, which were
adjusted as needed to ensure a given number of significant digits in
the solution.

\section*{Declaration of generative AI and AI-assisted technologies in the writing process}
During the preparation of this work, the author used an AI language
model to improve the readability and internal consistency of the text,
and for assistance with background research. After using this tool,
the author reviewed and edited the content as needed and takes full
responsibility for the content of the publication.
\bibliography{references}

\begin{thebibliography}{32}%
\makeatletter
\providecommand \@ifxundefined [1]{%
 \@ifx{#1\undefined}
}%
\providecommand \@ifnum [1]{%
 \ifnum #1\expandafter \@firstoftwo
 \else \expandafter \@secondoftwo
 \fi
}%
\providecommand \@ifx [1]{%
 \ifx #1\expandafter \@firstoftwo
 \else \expandafter \@secondoftwo
 \fi
}%
\providecommand \natexlab [1]{#1}%
\providecommand \enquote  [1]{``#1''}%
\providecommand \bibnamefont  [1]{#1}%
\providecommand \bibfnamefont [1]{#1}%
\providecommand \citenamefont [1]{#1}%
\providecommand \href@noop [0]{\@secondoftwo}%
\providecommand \href [0]{\begingroup \@sanitize@url \@href}%
\providecommand \@href[1]{\@@startlink{#1}\@@href}%
\providecommand \@@href[1]{\endgroup#1\@@endlink}%
\providecommand \@sanitize@url [0]{\catcode `\\12\catcode `\$12\catcode
  `\&12\catcode `\#12\catcode `\^12\catcode `\_12\catcode `\%12\relax}%
\providecommand \@@startlink[1]{}%
\providecommand \@@endlink[0]{}%
\providecommand \url  [0]{\begingroup\@sanitize@url \@url }%
\providecommand \@url [1]{\endgroup\@href {#1}{\urlprefix }}%
\providecommand \urlprefix  [0]{URL }%
\providecommand \Eprint [0]{\href }%
\providecommand \doibase [0]{https://doi.org/}%
\providecommand \selectlanguage [0]{\@gobble}%
\providecommand \bibinfo  [0]{\@secondoftwo}%
\providecommand \bibfield  [0]{\@secondoftwo}%
\providecommand \translation [1]{[#1]}%
\providecommand \BibitemOpen [0]{}%
\providecommand \bibitemStop [0]{}%
\providecommand \bibitemNoStop [0]{.\EOS\space}%
\providecommand \EOS [0]{\spacefactor3000\relax}%
\providecommand \BibitemShut  [1]{\csname bibitem#1\endcsname}%
\let\auto@bib@innerbib\@empty
\bibitem [{\citenamefont {Goldstein}\ \emph {et~al.}(2002)\citenamefont
  {Goldstein}, \citenamefont {Poole},\ and\ \citenamefont
  {Safko}}]{Goldstein2002}%
  \BibitemOpen
  \bibfield  {author} {\bibinfo {author} {\bibfnamefont {H.}~\bibnamefont
  {Goldstein}}, \bibinfo {author} {\bibfnamefont {C.~P.}\ \bibnamefont
  {Poole}},\ and\ \bibinfo {author} {\bibfnamefont {J.~L.}\ \bibnamefont
  {Safko}},\ }\href@noop {} {\emph {\bibinfo {title} {Classical Mechanics}}},\
  \bibinfo {edition} {3rd}\ ed.\ (\bibinfo  {publisher} {Addison Wesley},\
  \bibinfo {year} {2002})\BibitemShut {NoStop}%
\bibitem [{\citenamefont {Lorenz}(1963)}]{Lorenz1963}%
  \BibitemOpen
  \bibfield  {author} {\bibinfo {author} {\bibfnamefont {E.~N.}\ \bibnamefont
  {Lorenz}},\ }\bibfield  {title} {\bibinfo {title} {Deterministic nonperiodic
  flow},\ }\href@noop {} {\bibfield  {journal} {\bibinfo  {journal} {Journal of
  the Atmospheric Sciences}\ }\textbf {\bibinfo {volume} {20}},\ \bibinfo
  {pages} {130} (\bibinfo {year} {1963})}\BibitemShut {NoStop}%
\bibitem [{\citenamefont {Strogatz}(2015)}]{Strogatz2015}%
  \BibitemOpen
  \bibfield  {author} {\bibinfo {author} {\bibfnamefont {S.~H.}\ \bibnamefont
  {Strogatz}},\ }\href@noop {} {\emph {\bibinfo {title} {Nonlinear Dynamics and
  Chaos: With Applications to Physics, Biology, Chemistry, and Engineering}}},\
  \bibinfo {edition} {2nd}\ ed.\ (\bibinfo  {publisher} {Westview Press},\
  \bibinfo {year} {2015})\BibitemShut {NoStop}%
\bibitem [{\citenamefont {Tarasov}(2010)}]{Tarasov2010}%
  \BibitemOpen
  \bibfield  {author} {\bibinfo {author} {\bibfnamefont {V.~E.}\ \bibnamefont
  {Tarasov}},\ }\bibfield  {title} {\bibinfo {title} {Fractional dynamical
  systems},\ }in\ \href@noop {} {\emph {\bibinfo {booktitle} {Fractional
  Dynamics: Applications of Fractional Calculus to Dynamics of Particles,
  Fields and Media}}}\ (\bibinfo  {publisher} {Springer Berlin Heidelberg},\
  \bibinfo {address} {Berlin, Heidelberg},\ \bibinfo {year} {2010})\ pp.\
  \bibinfo {pages} {293--313}\BibitemShut {NoStop}%
\bibitem [{\citenamefont {Cresson}(2007)}]{Cresson2007}%
  \BibitemOpen
  \bibfield  {author} {\bibinfo {author} {\bibfnamefont {J.}~\bibnamefont
  {Cresson}},\ }\bibfield  {title} {\bibinfo {title} {Fractional embedding of
  differential operators and lagrangian systems},\ }\href@noop {} {\bibfield
  {journal} {\bibinfo  {journal} {Journal of Mathematical Physics}\ }\textbf
  {\bibinfo {volume} {48}},\ \bibinfo {pages} {033504} (\bibinfo {year}
  {2007})}\BibitemShut {NoStop}%
\bibitem [{\citenamefont {Palmer}(1993)}]{Palmer1993}%
  \BibitemOpen
  \bibfield  {author} {\bibinfo {author} {\bibfnamefont {T.~N.}\ \bibnamefont
  {Palmer}},\ }\bibfield  {title} {\bibinfo {title} {Extended-range atmospheric
  prediction and the lorenz model},\ }\href@noop {} {\bibfield  {journal}
  {\bibinfo  {journal} {Bulletin of the American Meteorological Society}\
  }\textbf {\bibinfo {volume} {74}},\ \bibinfo {pages} {49} (\bibinfo {year}
  {1993})}\BibitemShut {NoStop}%
\bibitem [{\citenamefont {Haken}(1975)}]{Haken1975}%
  \BibitemOpen
  \bibfield  {author} {\bibinfo {author} {\bibfnamefont {H.}~\bibnamefont
  {Haken}},\ }\bibfield  {title} {\bibinfo {title} {Analogy between higher
  instabilities in fluids and lasers},\ }\href@noop {} {\bibfield  {journal}
  {\bibinfo  {journal} {Physics Letters A}\ }\textbf {\bibinfo {volume} {53}},\
  \bibinfo {pages} {77} (\bibinfo {year} {1975})}\BibitemShut {NoStop}%
\bibitem [{\citenamefont {Pecora}\ and\ \citenamefont
  {Carroll}(1990)}]{Pecora1990}%
  \BibitemOpen
  \bibfield  {author} {\bibinfo {author} {\bibfnamefont {L.~M.}\ \bibnamefont
  {Pecora}}\ and\ \bibinfo {author} {\bibfnamefont {T.~L.}\ \bibnamefont
  {Carroll}},\ }\bibfield  {title} {\bibinfo {title} {Synchronization in
  chaotic systems},\ }\href@noop {} {\bibfield  {journal} {\bibinfo  {journal}
  {Phys. Rev. Lett.}\ }\textbf {\bibinfo {volume} {64}},\ \bibinfo {pages}
  {821} (\bibinfo {year} {1990})}\BibitemShut {NoStop}%
\bibitem [{\citenamefont {Liu}\ \emph {et~al.}(2022)\citenamefont {Liu},
  \citenamefont {Miao},\ and\ \citenamefont {Liu}}]{Liu2022}%
  \BibitemOpen
  \bibfield  {author} {\bibinfo {author} {\bibfnamefont {L.}~\bibnamefont
  {Liu}}, \bibinfo {author} {\bibfnamefont {S.}~\bibnamefont {Miao}},\ and\
  \bibinfo {author} {\bibfnamefont {S.}~\bibnamefont {Liu}},\ }\bibfield
  {title} {\bibinfo {title} {A novel image encryption algorithm based on lorenz
  system and dynamic s-box},\ }\href@noop {} {\bibfield  {journal} {\bibinfo
  {journal} {Cryptography}\ }\textbf {\bibinfo {volume} {6}},\ \bibinfo {pages}
  {53} (\bibinfo {year} {2022})}\BibitemShut {NoStop}%
\bibitem [{\citenamefont {Scott}\ \emph {et~al.}(1991)\citenamefont {Scott},
  \citenamefont {Peng}, \citenamefont {Tomlin},\ and\ \citenamefont
  {Showalter}}]{Scott1991}%
  \BibitemOpen
  \bibfield  {author} {\bibinfo {author} {\bibfnamefont {S.~K.}\ \bibnamefont
  {Scott}}, \bibinfo {author} {\bibfnamefont {B.}~\bibnamefont {Peng}},
  \bibinfo {author} {\bibfnamefont {A.~S.}\ \bibnamefont {Tomlin}},\ and\
  \bibinfo {author} {\bibfnamefont {K.}~\bibnamefont {Showalter}},\ }\bibfield
  {title} {\bibinfo {title} {Transient chaos in a closed chemical system},\
  }\href@noop {} {\bibfield  {journal} {\bibinfo  {journal} {The Journal of
  Chemical Physics}\ }\textbf {\bibinfo {volume} {94}},\ \bibinfo {pages}
  {1134} (\bibinfo {year} {1991})}\BibitemShut {NoStop}%
\bibitem [{\citenamefont {Schwartz}(2008)}]{Schwartz2008}%
  \BibitemOpen
  \bibfield  {author} {\bibinfo {author} {\bibfnamefont {R.}~\bibnamefont
  {Schwartz}},\ }\href@noop {} {\emph {\bibinfo {title} {Biological Modeling
  and Simulation: A Survey of Practical Methodologies}}}\ (\bibinfo
  {publisher} {MIT Press},\ \bibinfo {address} {Cambridge, MA},\ \bibinfo
  {year} {2008})\BibitemShut {NoStop}%
\bibitem [{\citenamefont {Mandelbrot}(1963)}]{Mandelbrot1963}%
  \BibitemOpen
  \bibfield  {author} {\bibinfo {author} {\bibfnamefont {B.}~\bibnamefont
  {Mandelbrot}},\ }\bibfield  {title} {\bibinfo {title} {The variation of
  certain speculative prices},\ }\href@noop {} {\bibfield  {journal} {\bibinfo
  {journal} {The Journal of Business}\ }\textbf {\bibinfo {volume} {36}},\
  \bibinfo {pages} {394} (\bibinfo {year} {1963})}\BibitemShut {NoStop}%
\bibitem [{\citenamefont {Ghys}(2007)}]{Ghys2007}%
  \BibitemOpen
  \bibfield  {author} {\bibinfo {author} {\bibfnamefont {E.}~\bibnamefont
  {Ghys}},\ }\bibfield  {title} {\bibinfo {title} {Knots and dynamics},\ }in\
  \href@noop {} {\emph {\bibinfo {booktitle} {Proceedings of the International
  Congress of Mathematicians, Vol. 1}}}\ (\bibinfo  {publisher} {European
  Mathematical Society},\ \bibinfo {address} {Madrid},\ \bibinfo {year}
  {2007})\ pp.\ \bibinfo {pages} {247--277}\BibitemShut {NoStop}%
\bibitem [{\citenamefont {Birman}\ and\ \citenamefont
  {Williams}(1983)}]{BirmanWilliams1983}%
  \BibitemOpen
  \bibfield  {author} {\bibinfo {author} {\bibfnamefont {J.~S.}\ \bibnamefont
  {Birman}}\ and\ \bibinfo {author} {\bibfnamefont {R.~F.}\ \bibnamefont
  {Williams}},\ }\bibfield  {title} {\bibinfo {title} {Knotted periodic orbits
  in dynamical systems—i: Lorenz’s equations},\ }\href@noop {} {\bibfield
  {journal} {\bibinfo  {journal} {Topology}\ }\textbf {\bibinfo {volume}
  {22}},\ \bibinfo {pages} {47} (\bibinfo {year} {1983})}\BibitemShut {NoStop}%
\bibitem [{\citenamefont {Ott}(2002)}]{Ott2002}%
  \BibitemOpen
  \bibfield  {author} {\bibinfo {author} {\bibfnamefont {E.}~\bibnamefont
  {Ott}},\ }\href@noop {} {\emph {\bibinfo {title} {Chaos in Dynamical
  Systems}}},\ \bibinfo {edition} {2nd}\ ed.\ (\bibinfo  {publisher} {Cambridge
  University Press},\ \bibinfo {year} {2002})\BibitemShut {NoStop}%
\bibitem [{\citenamefont {Alberti}\ \emph {et~al.}(2023)\citenamefont
  {Alberti}, \citenamefont {Faranda}, \citenamefont {Lucarini}, \citenamefont
  {Donner}, \citenamefont {Dubrulle},\ and\ \citenamefont
  {Daviaud}}]{Alberti2023}%
  \BibitemOpen
  \bibfield  {author} {\bibinfo {author} {\bibfnamefont {T.}~\bibnamefont
  {Alberti}}, \bibinfo {author} {\bibfnamefont {D.}~\bibnamefont {Faranda}},
  \bibinfo {author} {\bibfnamefont {V.}~\bibnamefont {Lucarini}}, \bibinfo
  {author} {\bibfnamefont {R.~V.}\ \bibnamefont {Donner}}, \bibinfo {author}
  {\bibfnamefont {B.}~\bibnamefont {Dubrulle}},\ and\ \bibinfo {author}
  {\bibfnamefont {F.}~\bibnamefont {Daviaud}},\ }\bibfield  {title} {\bibinfo
  {title} {Scale dependence of fractal dimension in deterministic and
  stochastic lorenz-63 systems},\ }\href@noop {} {\bibfield  {journal}
  {\bibinfo  {journal} {Chaos: An Interdisciplinary Journal of Nonlinear
  Science}\ }\textbf {\bibinfo {volume} {33}},\ \bibinfo {pages} {023144}
  (\bibinfo {year} {2023})}\BibitemShut {NoStop}%
\bibitem [{\citenamefont {Kuznetsov}\ \emph {et~al.}(2020)\citenamefont
  {Kuznetsov}, \citenamefont {Mokaev}, \citenamefont {Kuznetsova},\ and\
  \citenamefont {Kudryashova}}]{kuznetsov2020lorenz}%
  \BibitemOpen
  \bibfield  {author} {\bibinfo {author} {\bibfnamefont {N.}~\bibnamefont
  {Kuznetsov}}, \bibinfo {author} {\bibfnamefont {T.}~\bibnamefont {Mokaev}},
  \bibinfo {author} {\bibfnamefont {O.}~\bibnamefont {Kuznetsova}},\ and\
  \bibinfo {author} {\bibfnamefont {E.}~\bibnamefont {Kudryashova}},\
  }\bibfield  {title} {\bibinfo {title} {{The Lorenz system: hidden boundary of
  practical stability and the Lyapunov dimension}},\ }\href
  {https://doi.org/10.1007/s11071-020-05856-4} {\bibfield  {journal} {\bibinfo
  {journal} {Nonlinear Dynamics}\ }\textbf {\bibinfo {volume} {102}},\ \bibinfo
  {pages} {713} (\bibinfo {year} {2020})}\BibitemShut {NoStop}%
\bibitem [{\citenamefont {Sparrow}(1982)}]{Sparrow1982}%
  \BibitemOpen
  \bibfield  {author} {\bibinfo {author} {\bibfnamefont {C.}~\bibnamefont
  {Sparrow}},\ }\href@noop {} {\emph {\bibinfo {title} {The Lorenz Equations:
  Bifurcations, Chaos, and Strange Attractors}}}\ (\bibinfo  {publisher}
  {Springer-Verlag},\ \bibinfo {year} {1982})\BibitemShut {NoStop}%
\bibitem [{\citenamefont {Tabor}(1989)}]{Tabor1989}%
  \BibitemOpen
  \bibfield  {author} {\bibinfo {author} {\bibfnamefont {M.}~\bibnamefont
  {Tabor}},\ }\href@noop {} {\emph {\bibinfo {title} {Chaos and Integrability
  in Nonlinear Dynamics: An Introduction}}}\ (\bibinfo  {publisher} {Wiley},\
  \bibinfo {year} {1989})\BibitemShut {NoStop}%
\bibitem [{\citenamefont {Riewe}(1996)}]{Riewe1996}%
  \BibitemOpen
  \bibfield  {author} {\bibinfo {author} {\bibfnamefont {F.}~\bibnamefont
  {Riewe}},\ }\bibfield  {title} {\bibinfo {title} {Nonconservative lagrangian
  and hamiltonian mechanics},\ }\href@noop {} {\bibfield  {journal} {\bibinfo
  {journal} {Physical Review E}\ }\textbf {\bibinfo {volume} {53}},\ \bibinfo
  {pages} {1890} (\bibinfo {year} {1996})}\BibitemShut {NoStop}%
\bibitem [{\citenamefont {Musielak}(2009)}]{Musielak2008}%
  \BibitemOpen
  \bibfield  {author} {\bibinfo {author} {\bibfnamefont {Z.}~\bibnamefont
  {Musielak}},\ }\bibfield  {title} {\bibinfo {title} {General conditions for
  the existence of non-standard lagrangians for dissipative dynamical
  systems},\ }\href
  {https://doi.org/https://doi.org/10.1016/j.chaos.2009.03.171} {\bibfield
  {journal} {\bibinfo  {journal} {Chaos, Solitons \and Fractals}\ }\textbf
  {\bibinfo {volume} {42}},\ \bibinfo {pages} {2645} (\bibinfo {year}
  {2009})}\BibitemShut {NoStop}%
\bibitem [{\citenamefont {Auerbach}\ \emph {et~al.}(1987)\citenamefont
  {Auerbach}, \citenamefont {Cvitanovi{\'{c}}}, \citenamefont {Eckmann},
  \citenamefont {Gunaratne},\ and\ \citenamefont {Procaccia}}]{Auerbach1987}%
  \BibitemOpen
  \bibfield  {author} {\bibinfo {author} {\bibfnamefont {D.}~\bibnamefont
  {Auerbach}}, \bibinfo {author} {\bibfnamefont {P.}~\bibnamefont
  {Cvitanovi{\'{c}}}}, \bibinfo {author} {\bibfnamefont {J.-P.}\ \bibnamefont
  {Eckmann}}, \bibinfo {author} {\bibfnamefont {G.~H.}\ \bibnamefont
  {Gunaratne}},\ and\ \bibinfo {author} {\bibfnamefont {I.}~\bibnamefont
  {Procaccia}},\ }\bibfield  {title} {\bibinfo {title} {{Exploring Chaotic
  Motion Through Periodic Orbits}},\ }\href
  {https://doi.org/10.1103/PhysRevLett.58.2387} {\bibfield  {journal} {\bibinfo
   {journal} {Physical Review Letters}\ }\textbf {\bibinfo {volume} {58}},\
  \bibinfo {pages} {2387} (\bibinfo {year} {1987})}\BibitemShut {NoStop}%
\bibitem [{\citenamefont {Moon}\ \emph {et~al.}(2019)\citenamefont {Moon},
  \citenamefont {Seo}, \citenamefont {Han}, \citenamefont {Park},\ and\
  \citenamefont {Baik}}]{Moon2019}%
  \BibitemOpen
  \bibfield  {author} {\bibinfo {author} {\bibfnamefont {S.}~\bibnamefont
  {Moon}}, \bibinfo {author} {\bibfnamefont {J.~M.}\ \bibnamefont {Seo}},
  \bibinfo {author} {\bibfnamefont {B.-S.}\ \bibnamefont {Han}}, \bibinfo
  {author} {\bibfnamefont {J.}~\bibnamefont {Park}},\ and\ \bibinfo {author}
  {\bibfnamefont {J.-J.}\ \bibnamefont {Baik}},\ }\bibfield  {title} {\bibinfo
  {title} {A physically extended {L}orenz system},\ }\href
  {https://doi.org/10.1063/1.5095466} {\bibfield  {journal} {\bibinfo
  {journal} {Chaos: An Interdisciplinary Journal of Nonlinear Science}\
  }\textbf {\bibinfo {volume} {29}},\ \bibinfo {pages} {063129} (\bibinfo
  {year} {2019})}\BibitemShut {NoStop}%
\bibitem [{\citenamefont {Park}\ \emph {et~al.}(2021)\citenamefont {Park},
  \citenamefont {Moon}, \citenamefont {Seo},\ and\ \citenamefont
  {Baik}}]{Park2021}%
  \BibitemOpen
  \bibfield  {author} {\bibinfo {author} {\bibfnamefont {J.}~\bibnamefont
  {Park}}, \bibinfo {author} {\bibfnamefont {S.}~\bibnamefont {Moon}}, \bibinfo
  {author} {\bibfnamefont {J.~M.}\ \bibnamefont {Seo}},\ and\ \bibinfo {author}
  {\bibfnamefont {J.-J.}\ \bibnamefont {Baik}},\ }\bibfield  {title} {\bibinfo
  {title} {Systematic comparison between the generalized {L}orenz equations and
  {DNS} in the two-dimensional {R}ayleigh--{B}énard convection},\ }\href
  {https://doi.org/10.1063/5.0051482} {\bibfield  {journal} {\bibinfo
  {journal} {Chaos: An Interdisciplinary Journal of Nonlinear Science}\
  }\textbf {\bibinfo {volume} {31}},\ \bibinfo {pages} {073119} (\bibinfo
  {year} {2021})}\BibitemShut {NoStop}%
\bibitem [{\citenamefont {Cvitanovi\'c}\ \emph {et~al.}(2005)\citenamefont
  {Cvitanovi\'c} \emph {et~al.}}]{Cvitanovic2005}%
  \BibitemOpen
  \bibfield  {author} {\bibinfo {author} {\bibfnamefont {P.}~\bibnamefont
  {Cvitanovi\'c}} \emph {et~al.},\ }\href@noop {} {\emph {\bibinfo {title}
  {Chaos: Classical and Quantum}}}\ (\bibinfo  {publisher} {Niels Bohr
  Institute},\ \bibinfo {year} {2005})\ \bibinfo {note} {online version
  available at \url{https://chaosbook.org}}\BibitemShut {NoStop}%
\bibitem [{\citenamefont {Noether}(1918)}]{Noether1918}%
  \BibitemOpen
  \bibfield  {author} {\bibinfo {author} {\bibfnamefont {E.}~\bibnamefont
  {Noether}},\ }\bibfield  {title} {\bibinfo {title} {Invariante
  variationsprobleme},\ }\href@noop {} {\bibfield  {journal} {\bibinfo
  {journal} {Nachrichten von der Gesellschaft der Wissenschaften zu
  G\"ottingen, Mathematisch-Physikalische Klasse}\ ,\ \bibinfo {pages} {235}}
  (\bibinfo {year} {1918})},\ \bibinfo {note} {english translation: Transport
  Theory and Statistical Physics, 1(3), 186--207 (1971)}\BibitemShut {NoStop}%
\bibitem [{\citenamefont {Tucker}(2002)}]{Tucker2002}%
  \BibitemOpen
  \bibfield  {author} {\bibinfo {author} {\bibfnamefont {W.}~\bibnamefont
  {Tucker}},\ }\bibfield  {title} {\bibinfo {title} {A rigorous ode solver and
  smale’s 14th problem},\ }\href@noop {} {\bibfield  {journal} {\bibinfo
  {journal} {Foundations of Computational Mathematics}\ }\textbf {\bibinfo
  {volume} {2}},\ \bibinfo {pages} {53} (\bibinfo {year} {2002})}\BibitemShut
  {NoStop}%
\bibitem [{\citenamefont {Angelis}\ \emph {et~al.}(2023)\citenamefont
  {Angelis}, \citenamefont {Sofos},\ and\ \citenamefont
  {Karakasidis}}]{Angelis2023}%
  \BibitemOpen
  \bibfield  {author} {\bibinfo {author} {\bibfnamefont {D.}~\bibnamefont
  {Angelis}}, \bibinfo {author} {\bibfnamefont {F.}~\bibnamefont {Sofos}},\
  and\ \bibinfo {author} {\bibfnamefont {T.~E.}\ \bibnamefont {Karakasidis}},\
  }\bibfield  {title} {\bibinfo {title} {Artificial intelligence in physical
  sciences: Symbolic regression trends and perspectives},\ }\href@noop {}
  {\bibfield  {journal} {\bibinfo  {journal} {Archives of Computational Methods
  in Engineering}\ }\textbf {\bibinfo {volume} {30}},\ \bibinfo {pages} {3845}
  (\bibinfo {year} {2023})}\BibitemShut {NoStop}%
\bibitem [{\citenamefont {Schmidt}\ and\ \citenamefont
  {Lipson}(2009)}]{Schmidt2009}%
  \BibitemOpen
  \bibfield  {author} {\bibinfo {author} {\bibfnamefont {M.}~\bibnamefont
  {Schmidt}}\ and\ \bibinfo {author} {\bibfnamefont {H.}~\bibnamefont
  {Lipson}},\ }\bibfield  {title} {\bibinfo {title} {Distilling free-form
  natural laws from experimental data},\ }\href@noop {} {\bibfield  {journal}
  {\bibinfo  {journal} {Science}\ }\textbf {\bibinfo {volume} {324}},\ \bibinfo
  {pages} {81} (\bibinfo {year} {2009})}\BibitemShut {NoStop}%
\bibitem [{\citenamefont {Cava}\ \emph {et~al.}(2021)\citenamefont {Cava},
  \citenamefont {Orzechowski}, \citenamefont {Burlacu}, \citenamefont
  {Francca}, \citenamefont {Virgolin}, \citenamefont {Jin}, \citenamefont
  {Kommenda},\ and\ \citenamefont {Moore}}]{Cava2021}%
  \BibitemOpen
  \bibfield  {author} {\bibinfo {author} {\bibfnamefont {W.~L.}\ \bibnamefont
  {Cava}}, \bibinfo {author} {\bibfnamefont {P.}~\bibnamefont {Orzechowski}},
  \bibinfo {author} {\bibfnamefont {B.}~\bibnamefont {Burlacu}}, \bibinfo
  {author} {\bibfnamefont {F.~O.~d.}\ \bibnamefont {Francca}}, \bibinfo
  {author} {\bibfnamefont {M.}~\bibnamefont {Virgolin}}, \bibinfo {author}
  {\bibfnamefont {Y.}~\bibnamefont {Jin}}, \bibinfo {author} {\bibfnamefont
  {M.}~\bibnamefont {Kommenda}},\ and\ \bibinfo {author} {\bibfnamefont
  {J.~H.}\ \bibnamefont {Moore}},\ }\bibfield  {title} {\bibinfo {title}
  {Contemporary symbolic regression methods and their relative performance},\
  }\href@noop {} {\bibfield  {journal} {\bibinfo  {journal} {Advances in neural
  information processing systems 2021 DB1}\ ,\ \bibinfo {pages} {1}} (\bibinfo
  {year} {2021})}\BibitemShut {NoStop}%
\bibitem [{\citenamefont {Krishnamurthy}\ \emph {et~al.}(1999)\citenamefont
  {Krishnamurthy}, \citenamefont {Kinter~III},\ and\ \citenamefont
  {Shukla}}]{Krishnamurthy1999}%
  \BibitemOpen
  \bibfield  {author} {\bibinfo {author} {\bibfnamefont {V.}~\bibnamefont
  {Krishnamurthy}}, \bibinfo {author} {\bibfnamefont {J.~L.}\ \bibnamefont
  {Kinter~III}},\ and\ \bibinfo {author} {\bibfnamefont {J.}~\bibnamefont
  {Shukla}},\ }\bibfield  {title} {\bibinfo {title} {A physically extended
  lorenz system},\ }\href@noop {} {\bibfield  {journal} {\bibinfo  {journal}
  {Journal of the Atmospheric Sciences}\ }\textbf {\bibinfo {volume} {56}},\
  \bibinfo {pages} {29} (\bibinfo {year} {1999})}\BibitemShut {NoStop}%
\bibitem [{\citenamefont {Miranda}\ \emph {et~al.}(2013)\citenamefont
  {Miranda}, \citenamefont {Rempel}, \citenamefont {Chian}, \citenamefont
  {Seehafer}, \citenamefont {Toledo},\ and\ \citenamefont
  {Muñoz}}]{Miranda2013}%
  \BibitemOpen
  \bibfield  {author} {\bibinfo {author} {\bibfnamefont {R.~A.}\ \bibnamefont
  {Miranda}}, \bibinfo {author} {\bibfnamefont {E.~L.}\ \bibnamefont {Rempel}},
  \bibinfo {author} {\bibfnamefont {A.~C.-L.}\ \bibnamefont {Chian}}, \bibinfo
  {author} {\bibfnamefont {N.}~\bibnamefont {Seehafer}}, \bibinfo {author}
  {\bibfnamefont {B.~A.}\ \bibnamefont {Toledo}},\ and\ \bibinfo {author}
  {\bibfnamefont {P.~R.}\ \bibnamefont {Muñoz}},\ }\bibfield  {title}
  {\bibinfo {title} {Lagrangian coherent structures at the onset of hyperchaos
  in the two-dimensional navier-stokes equations},\ }\href
  {https://doi.org/10.1063/1.4811297} {\bibfield  {journal} {\bibinfo
  {journal} {Chaos: An Interdisciplinary Journal of Nonlinear Science}\
  }\textbf {\bibinfo {volume} {23}},\ \bibinfo {pages} {033107} (\bibinfo
  {year} {2013})}\BibitemShut {NoStop}%
\end{thebibliography}%

\end{document}